\documentclass[research]{letinbiom} 

\usepackage{natbib}
\usepackage{bm}
\title{SEIRD MODEL FOR QATAR COVID-19 OUTBREAK: A CASE STUDY
}

\author{%
  Ryad Ghanam\rlap,\inst{a}
  Edward~L.\ Boone \rlap,\inst{b}
  Abdel-Salam~G.\ Abdel-Salam\inst{c}
}

\institution{
  \inst{a}Department of Liberal Arts and Sciences, Virginia Commonwealth University in Qatar, Education City, Doha, Qatar.;
  \inst{b}Department of Statistical Sciences and Operations Research, Virginia Commonwealth University, Richmond, Virginia, 23284, USA;
  \inst{c}Department of Mathematics, Statistics and Physics,
College of Arts and Sciences, Qatar University, Doha, Qatar}

\correspondence{Edward~L.\ Boone}{elboone@vcu.edu}

\authorheading{R.~Ghanam, E.~Boone, A-S.\,G.~Abdel-Salam}


\abstract{%
The Covid-19 outbreak of 2020 has required many governments to develop mathematical-statistical models of the outbreak for policy and planning purposes.  This work provides a tutorial on building a compartmental model using Susceptibles, Exposed, Infected, Recovered and Deaths status through time.  A Bayesian Framework is utilized to perform both parameter estimation and predictions.  This model uses interventions to quantify the impact of various government attempts to slow the spread of the virus.  Predictions are also made to determine when the peak Active Infections will occur.
}

\begin{document}

\maketitle


\section{Introduction}	

Coronavirus Disease 2019 (COVID-19) (\citet{Wu};  \citet{Rezabakhsh}) is a severe pandemic affecting the whole world with a fast spreading regime, requiring to perform strict precautions to keep it under control. As there is no cure and target treatment yet, establishing those precautions become inevitable.  These limitations (\citet{Giuliani}) can be listed as social distancing, closure of businesses and schools and travel prohibitions (\citet{Chinazzi}).

Corona Virus is a new human Betacoronavirus that uses densely glycosylated spike protein to penetrate host cells. The COVID-19 belongs to the same family classification with Nidovirales, viruses that use a nested set of mRNAs to replicate and it further falls under the subfamily of alpha, beta, gamma and delta Co-Vis. The virus that causes COVID-19 belongs to the Betacoronavirus 2B lineage and has a close relationship with SARS species. It is a novel virus since the monoclonal antibodies do not exhibit a high degree of binding to SARS-CoV-2. Replication of the viral RNA occurs when RNA polymerase binds and re-attaches to multiple locations (\citet{McIntosh}; \citet{Fisher}).

Cases of COVID-19 started in December 2019 when a strange condition was reported in Wuhan, China. This virus has a global mortality rate of 3.4\%, which makes it more severe in relation to flu. The elderly who have other pre-existing illnesses are succumbing more to the COVID-19. People with only mild symptoms recover within 3 to 7 days, while those with conditions such as pneumonia or severe diseases take weeks to recover. The recovery percentage of patients, for example, in China stands at 51\%.  The recovery percentage rate of COVID-19 is expected to hit 90\% (\citet{World}).

The virus has spread from China to 196 other countries and territories across the globe. From Wuhan, Hubei province, the virus spread to Mainland China, Thailand, Japan, South Korea, Vietnam, Singapore, Italy, Iran, and other countries. The State of Qatar was one of the countries that were affected by the COVID-19 spreading and the first infected case was reported on $29^{th}$ of February 2020 and it could be considered the $2^{nd}$ highest in the Arab World with the number of confirmed cases $28,272$ as of May 14, 2020.

For effectively specifying such security measures, it is essential to have a real-time monitoring system of the infection, recovery and death rates. Develop, implement and deploy a data-driven forecasting model for use by stakeholders in the State of Qatar to deal with the Covid-19 pandemic.  The model will focus on infected, deaths and recovered as those are the only data available at this time.

This document is organized in the following manner.  In Section~\ref{sec:SEIRD} the SEIRD that is employed is defined.  Next, Section~\ref{sec:Data} introduces the data available and gives description.  Then \ref{sec:Inter} shows how interventions are incorporated into the model.  The Bayesian inference model specification is given in Section~\ref{sec:Bayes}.  Summaries of the parameter estimation is shown in Section~\ref{sec:Results}.  Predictive performance is provided in Section~\ref{sec:PP}.  And finally, a discussion is found in Section~\ref{sec:Discussion}.


\section{The SEIRD Model}\label{sec:SEIRD}

Let $S(t)$ be the number of people Susceptible at time $t$, $E(t)$ be the number of people Exposed at time t, $I(t)$ be the number of Infected at time $t$, $R(t)$ be the cumulative number of recovered at time $t$ and $D(t)$ be the cumulative number of Deaths at time $t$.  This can be modeled with the following system of ordinary differential equations:

\begin{eqnarray}\label{eq:Sys1}
\frac{\partial S(t)}{\partial t} &=& -\alpha S(t)E(t)    \cr
\frac{\partial E(t)}{\partial t} &=& \alpha S(t)E(t) - \beta E(t) \cr
\frac{\partial I(t)}{\partial t} &=& \beta E(t) - \gamma I(t) - \eta I(t)  \cr
\frac{\partial R(t)}{\partial t} &=& \gamma I(t) \cr
\frac{\partial D(t)}{\partial t} &=& \eta I(t)
\end{eqnarray}

\noindent where $\alpha$ is the transmission rate from Susceptibles to Exposed, $\beta$ is the rate at which Exposed become Infected, $\gamma$ is the rate at which Infected become recovered and $\eta$ is the mortality rate for those Infected.  Notice that, this model formulation makes several key assumptions:
\begin{enumerate}
  \item Immigration, emigration, natural mortality and births are negligible over the time frame and hence are not in the model.
  \item Once a person is in the Infected group, they are quarantined and hence they do not mix with the Susceptible population.
  \item The Recovered and Deaths compartments are for those who first are infected.  There is no compartment for those Exposed who do not become sick (Infected) and Recover on their own.
\end{enumerate}

Traditional analysis would include a steady state analysis, however, in this case the dynamics of the short term is of interest.  Hence, this work does not address any steady state or equilibrium concerns.  This work is concerned with fitting the model given in (\ref{eq:Sys1}) to the Covid-19 data concerning the State of Qatar during the 2020 outbreak and using the model for forecasting several possible scenarios.

\section{Data}\label{sec:Data}

The Johns Hopkins Covid-19 github site includes for every country for each day the cumulative number of confirmed infections, cumulative number of recovered and the cumulative number of deaths for each day starting 22 January 2020.  The data for Qatar was obtained.  Notice that in model (\ref{eq:Sys1}) the Recovered and Death states are cumulative as once one enters the compartment their is no exit.  However, the Infected compartment has transitions from Exposed and to Recovered and Deaths.  Hence the data provided for confirmed infections is cumulative and included both Recovered and Deaths and will need to be removed from this compartment's data.  Let $CI(t)$ be the Confirmed Infections at time $t$ and let Infected $I(t)$ be defined as:
\begin{equation}
I(t) = CI(t) - R(t) - D(t)
\end{equation}
\noindent For clarity the term ``Active Infections'' will be used to denote this derived variable versus the cumulative Infected provided in the data.

\begin{figure}[ht!]
\begin{center}
  \begin{tabular}{ccc}
  (a) & (b) & (c) \cr
  \includegraphics[width = 0.3\textwidth]{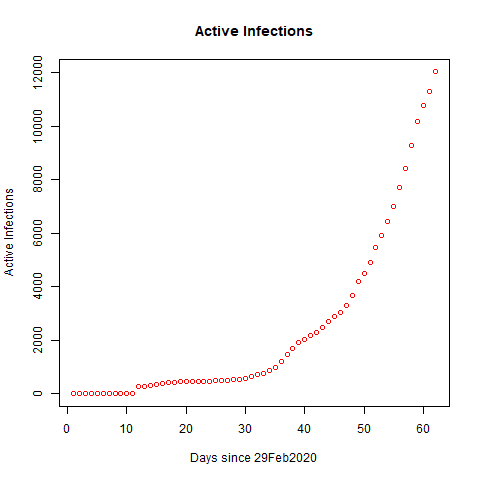} &
  \includegraphics[width = 0.3\textwidth]{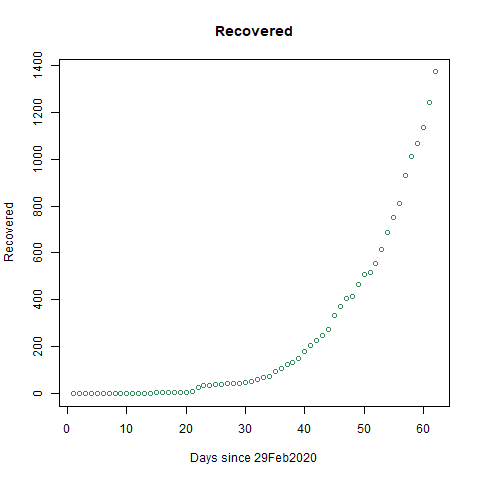} &
  \includegraphics[width = 0.3\textwidth]{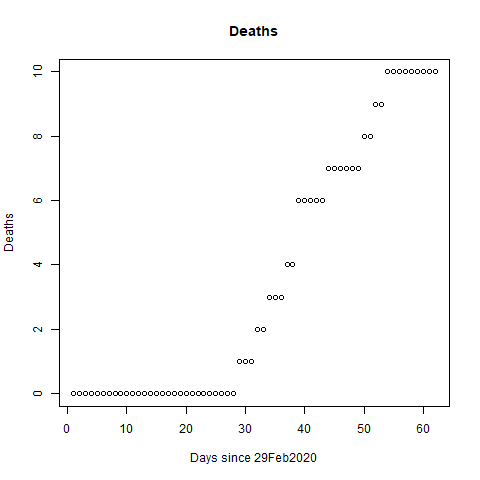}  \cr
  \end{tabular}\caption{Plots of Active Infections (a), Recovered (b) and Deaths (c) for the State of Qatar for the days since 29 February 2020 until 1 May 2020. \label{fig:InitialData}}
\end{center}
\end{figure}

Figure~\ref{fig:InitialData} shows the plots of the Active Infections, Recovered and Deaths data for Qatar for the days since 29 February 2020.  Notice that the Active Infections are very low until around day 12 when there is large jump due to increased testing.  The Active Infections then seems to plateau for until day 30, after which there is extreme growth in Active Infections.  There seems to be a similar pattern for the Recovered with a delay showing the time of infection before recovery.  The plot for Deaths shows no deaths until day 30 and then a steady increase in Deaths for the remaining days.

The State of Qatar, prepared an excellent flexible plan for risk management, grounded on national risk assessment, taking account of the global risk assessment done by WHO, focuses on reinforce capacities to reduce or eliminate the health risks from COVID-19. Embed complete emergency risk management strategy in the health sector. Furthermore, Enabling and promoting multi-sectoral linkage and integration across the whole-of-government and the whole-of-society(\citep{DoH}, \citep{DoH2}, and \citep{WHO}).

On March 9, 2020 (day 10), Qatar announced a closure of all universities and schools.  It placed a travel ban on 15 countries: Bangladesh, China, Egypt, India, Iran, Iraq, Italy, Lebanon, Nepal, Pakistan, the Philippines, South Korea, SriLanka, Syria, and Thailand. On March 14, 2020 (day 15), Qatar expanded its travel ban to include three new countries: Germany, Spain and France (\citet{HMC} and \citet{MPH}).

The Ministry of Municipal and Environment on March 21, 2020, closed all parks and public beaches to curb the spread of coronavirus.  On March 23, 2020 (day 24), the Ministry of Commerce and Industry decided to temporarily close all restaurants, cafes, food outlets, and food trucks at the main public era. Also, the Ministry of Commerce and Industry decided to close all the unnecessary business on March 27, 2020 (day 28) \citet{HMC} and \citet{MPH}.

These interventions taken by the government change the dynamics of the system and hence need to be incorporated into the model.  The next section details how we introduce interventions both from the government and interventions guided by the data.

\section{Interventions}\label{sec:Inter}
In Figure~\ref{fig:InitialData}, one can see the jump at day 12 and a plateau until day 30.  The model needs to be able to handle interventions made by the Government of the State of Qatar.  The main parameter that policy can influence is $\alpha$, the rate of transmission from Exposed to Susceptible.  One way to implement this the use of indicator functions $W_k(t)$ defined as:
\begin{equation}
W_k(t) = \begin{cases} 1 & \text{if } t > t_k \cr
0 & \text{otherwise}
\end{cases}
\end{equation}
\noindent where $t_k$ is the time where the $k^{th}$ intervention is taken and index $k = 1,2,..,K$.  For each intervention there needs to be a change to the value of $\alpha$, denoted $\alpha_k$, that captures the impact of the intervention.  Let the vector $W(t) = \left( 1, W_1(t), W_2(t), ... , W_K(t) \right)^{T}$ be the vector of the values of each $W_k(t)$ at time $t$.  Let $\bm{\alpha} = \left( \alpha_0, \alpha_1,...\alpha_K \right)^{T}$.  This formulation gives the following transitions rates between $S(t)$ and $E(t)$:
\begin{equation}
\alpha(t) = \begin{cases} \alpha_0 & \text{if } 0 < t < t_1 \cr
\alpha_0 + \alpha_1 & \text{if } t_1 < t < t_2 \cr
\alpha_0 + \alpha_1 + \alpha_2 & \text{if } t_2 < t < t_3 \cr
\vdots & \vdots \cr
\alpha_0 + \alpha_1 + ... \alpha_K & \text{if } t_{K} < t
\end{cases}
\end{equation}
\noindent which will require the following constraints due to the fact that $\alpha(t) > 0$ for all $t$:
\begin{eqnarray}
\alpha_0 &>& 0   \cr
\alpha_0 + \alpha_1 &>& 0  \cr
\alpha_0 + \alpha_1 + \alpha_2 &>& 0  \cr
\vdots &>& \vdots  \cr
\alpha_0 + \alpha_1 + ... +\alpha_K &>& 0
\end{eqnarray}
\noindent Let $\mathcal{A}$ be the set defined by the constraints above.

In addition to changes in infection rates $\bm{\alpha}$, impulse functions can be used to model dramatic one time shifts in transitions between states.  A Dirac delta function defined by 
\begin{equation}
\delta (x) = \begin{cases} +\infty, & \text{if } x=0 \cr
                             0, & \text{if } x\neq 0 
\end{cases}
\end{equation}
\noindent which satisfies $\int_{-\infty}^{\infty} \delta_(x) dx = 1$ \citep{Dirac}.  This can be integrated in the model to capture spikes in the number of cases.  In our case the State of Qatar data shows exhibits this type of behavior at day 12 where one can clearly see a large jump in the number of infections.  This is incorporated into the model presented by a Dirac delta function, $\delta(t-\tau)$, in transition rate between Exposed and Infected, which is coupled with a coefficient to $\beta_A$ to capture the impact of the jump.

\section{Bayesian Analysis}\label{sec:Bayes}
Due to the complexity of the model the Bayesian inferential framework is chosen.  Recall, Bayes formula is given by \citep{Bayes}:

\begin{equation}\
\pi (\theta | \mathbf{D} ) = \frac{ \pi( \theta )L( \mathbf{D} | \theta ) }{ \int_{\Theta} \pi( \theta )L( \mathbf{D} | \theta ) d\theta} \nonumber
\end{equation}
\noindent where $\pi( \theta | \mathbf{D})$ is the posterior probability distribution for the parameters $\theta$ given the data $\mathbf{D}$, $\pi(\theta)$ is the prior distribution of $\theta$ and $L( \mathbf{D} | \theta )$ is the likelihood of the data given $\theta$.

In order to specify the likelihood of the model in equation (\ref{eq:Sys1}) the model modified to model the {\it mean} abundance in each compartment and is given by:

\begin{eqnarray}\label{eq:Sys2}
\frac{\partial \lambda_S(t)}{\partial t} &=& -W(t)^T\bm{ \alpha } \lambda_S(t)\lambda_E(t)    \cr
\frac{\partial \lambda_E(t)}{\partial t} &=& W(t)^T\bm{ \alpha } \lambda_S(t)\lambda_E(t) - \beta \lambda_E(t) - \beta_A \lambda_E(t) \delta(t-\tau) \cr
\frac{\partial \lambda_I(t)}{\partial t} &=& \beta \lambda_E(t) + \beta_A \lambda_E(t)\delta(t-\tau) - \gamma \lambda_I(t) - \eta \lambda_I(t)  \cr
\frac{\partial \lambda_R(t)}{\partial t} &=& \gamma \lambda_I(t) \cr
\frac{\partial \lambda_D(t)}{\partial t} &=& \eta \lambda_I(t)
\end{eqnarray}

\noindent where $\lambda_S(t)$, $\lambda_E(t)$, $\lambda_I(t)$, $\lambda_R(t)$ and $\lambda_D(t)$ are the means of $S(t)$, $E(t)$, $I(t)$, $R(t)$ and $D(t)$, respectively and the parameters have the same definition as provided in the system given in equation (\ref{eq:Sys1}).  Since there is no data for $S(t)$ and $E(t)$ these compartments will be latent variables and will not directly factor into the likelihood.  The likelihood for $I(t)$, $R(t)$ and $D(t)$ are given by:

\begin{eqnarray}\label{eq:Like1}
I(t) &\sim& Poisson \left( \lambda_I(t) \right) \cr
R(t) &\sim& Poisson \left( \lambda_R(t) \right) \cr
D(t) &\sim& Poisson \left( \lambda_D(t) \right)
\end{eqnarray}

To specify the prior distributions for $\bm{ \alpha }$, $\beta_A$, $\beta$, $\gamma$ and $\eta$ one must incorporate the following constraints $\alpha > 0$, $\beta > 0$, $\gamma > 0$ and $\eta > 0$.  Hence the following prior distributions are set:
\begin{eqnarray}
\bm{ \alpha } &\sim& MVN(a\mathbf{1},\sigma^2\mathbf{I})C(\mathcal{A}) \cr
\beta_A &\sim& Exp(1) \cr
\beta &\sim& Exp(1) \cr
\gamma &\sim& Exp(1) \cr
\eta &\sim& Exp(1)
\end{eqnarray}
\noindent where $C(\mathcal{A})$ is an indicator function takes the value $1$ if $\bm{\alpha} \in \mathcal{A}$.  This serves to truncate the normal distribution in order to keep $\bm{\alpha}$ in the feasible range of values.

The likelihood and prior distributions specifications lead to the following posterior distribution when $a=1$ and $\sigma^2=1$:
\begin{eqnarray}
\pi \left( \bm{ \alpha }, \beta_A, \beta, \gamma, \eta | \mathbf{D} \right) &\propto& \pi(\bm{ \alpha })\pi(\beta)\pi(\gamma)\pi(\eta)L(\mathbf{D}|\bm{ \alpha }, \beta_A, \beta,\gamma,\eta) \cr
&=& C(\mathcal{A}) e^{-\frac{1}{2}(\bm{\alpha} - \mathbf{1})^T(\bm{\alpha} - \mathbf{1}) -\beta_A - \beta -\gamma-\eta}    \cr
&\times& \prod_{t=1}^{T}\frac{\lambda_I(t)^{I(t)}\lambda_R(t)^{R(t)}\lambda_D(t)^{D(t)}e^{-\lambda_I(t)-\lambda_R(t) - \lambda_D(t)}}{I(t)!R(t)!D(t)!}
\end{eqnarray}
The posterior distribution does not lend to any analytic solution, hence Markov chain Monte Carlo (MCMC) techniques will be used to sample from the posterior distribution \citep{Gelman}.  Specifically Metropolis-Hastings sampler is used to obtain samples from the posterior distribution \citep{Gilks} and \citep{Albert}.  To tune the sampler a series of short chains were generated and analyzed for convergence and adequate acceptance rates.  These initial short chains were discarded as ``burn-in'' samples.  The tuned sampler was used to generate 5,000 samples from $\pi( \bm{ \alpha }, \beta_A, \beta, \gamma, \eta | \mathbf{D})$ and trace plots were visually examined for convergence and deemed to be acceptable.  All inferences will be made from these 5,000 samples.  The model and sampling algorithm is custom programmed in the R statistical programming language version 3.6.3. The computation takes approximately 290 seconds using a AMD A10-9700 3.50GHz processor with 16GB of RAM to obtain 5,000 samples from the posterior distribution.  For more on statistical inference see \citet{MWS}, \citet{CasellaBerger}, and \citet{Berger}.

\section{Results}\label{sec:Results}
To apply the model the following initial conditions are specified: $S(0) = 2,782,000$, $E(0) = 3$, $I(0) = 1$, $R(0) = 0$ and $D(0) = 0$.  Here $S(0)$ is the current population of the State of Qatar, $I(0),~R(0)$ and $D(0)$ are obtained directly from the data.  The choice of $E(0)$ was used as it a minimal value that would allow the disease to spread but not so large as to make the spread rapid.  Several values of $E(0)$ were explored and the value of $3$ was found to have the best fit.  Furthermore, model interventions were placed at days $t_1 = 12,~t_2 = 24,~t_3 = 28,~t_4 = 40$ and $t_5 = 59$ with an Dirac delta impulse at time $\tau = 12$.

Table~\ref{tbl:ParEst} shows the means, standard deviations and the 0.025\%, 0.5\% and 0.975\% quantiles for the model parameters based on the 5,000 samples from the posterior distribution.  Notice that, $\alpha_0 = 2.33\times 10^{-7}$ and $\alpha_1 = -2.11\times 10^{-7}$ are very close in magnitude with different signs indicating that the first intervention drastically reduced the transmission rate.  Similarly one can see that the second and third interventions $\alpha_2 = 1.92\times 10^{-7}$ and $\alpha_3 = -1.74\times 10^{-7}$ essentially are of the same magnitude with different signs which when added resulting in a very low transmission rate.  However, $\alpha_4 = 1.83 \times 10^{-9}$ is a small increase with a moderate decrease in $\alpha_5 = -3.89 \times 10^{-8}$ which still leaves a final transmission rate of $\sum_{k=0}^K \alpha_k \approx 2.53\times10^{-9}$.  Of particular note is the mean mortality rate $\eta =0.00014 \approx 1/7142$ which means that about 1 in 7,142 people perish from the disease each day, which is quite low.  Also note that the mean infection (confirmed) rate is $\beta = 0.02818 \approx 1/35.48$ which corresponds to about 1 in 35.48 exposed people become confirmed each day.  The quantile intervals provide a 95\% credible interval for the parameters and can be used to obtain a range of reasonable parameter values.  For example for the parameter $\beta$ the interval is  (0.02478, 0.03298) meaning that the probability that $\beta$ is between (0.02478, 0.03298) is 0.95.  This can be used to create an interval for the risk interpretations as between $1/0.02478 \approx 40.35$ and $1/0.03298 \approx 30.32$ Exposed people are confirmed as infected each day.  This also gives insight into how many people who may be in the population who are Exposed and may be infectious but do not yet exhibit symptoms.  Recall that $\beta_A$ is associated with the Dirac delta function for impulse to model the jump in transition rate from Exposed to Infected at day 12.  Notice that $\beta_A \approx 0.79695$ means that there is a one time influx of approximately 79.6\% of the people Exposed moved to the Infected compartment. Hence the increased testing captured many of the Exposed people.  By adding this to the natural Exposed to Infected rate of $\beta = 0.02818$ one obtains the one time transmission rate of $\beta + \beta_A = 0.02818 + 0.79695 = 0.82513$ corresponding to a total of approximately 82.5\% of Exposed being confirmed as Infected.  Leaving the remaining approximately 17.5\% of Exposed people still interacting with the Susceptible population.

\begin{table}[ht!]
\caption{Mean, Standard Deviation and ($Q_{0.025},Q_{0.5},Q_{0.975}$) for $\alpha_0$, $\alpha_1$, $\alpha_2$, $\alpha_3$, $\alpha_4$, $\beta_A$, $\beta$, $\gamma$ and $\eta$.  Based on 5,000 samples from the posterior distribution}\label{tbl:ParEst}
\begin{center}
\begin{tabular}{lccc} \hline
Parameter & Mean & Std Dev.  & ($Q_{0.025},Q_{0.5},Q_{0.975}$)  \\ \hline
$\alpha_0$ &  2.33$\times 10^{-7}$ & 4.86$\times 10^{-10}$ &
(2.32$\times 10^{-7}$, 2.33$\times 10^{-7}$, 2.34$\times 10^{-7}$) \\
$\alpha_1$ (day 12) & -2.12$\times 10^{-7}$  & 4.70$\times 10^{-10}$ &
(-2.13$\times 10^{-7}$, -2.12$\times 10^{-7}$,-2.11$\times 10^{-7}$) \\
$\alpha_2$ (day 24) & 1.92$\times 10^{-7}$ & 4.53$\times 10^{-10}$ &
(1.91$\times 10^{-7}$,  1.92$\times 10^{-7}$,1.93$\times 10^{-7}$) \\
$\alpha_3$ (day 28) & -1.74$\times 10^{-7}$ & 2.06$\times 10^{-9}$ &
(-1.77$\times 10^{-7}$, -1.74$\times 10^{-7}$,-1.70$\times 10^{-7}$) \\
$\alpha_4$ (day 49) & 1.83$\times 10^{-9}$ & 8.48$\times 10^{-10}$ &
(3.35$\times 10^{-10}$, 1.75$\times 10^{-9}$, 3.61$\times 10^{-9}$) \\
$\alpha_5$ (day 59) & -3.89$\times 10^{-8}$ & 2.02$\times 10^{-9}$ &
(-4.19$\times 10^{-8}$, -3.93$\times 10^{-8}$, -3.39$\times 10^{-8}$) \\
$\beta_A$ (day 12) & 0.79695 & 0.00980 & (0.77516, 0.79781, 0.81379)  \\
$\beta$ & 0.02818 & 0.00263 & (0.02478, 0.02760, 0.03298)  \\
$\gamma$ & 0.00980 & 9.51$\times 10^{-5}$ & (0.00956, 0.00974, 0.00991)  \\
$\eta$ & 0.00014 & 8.49$\times 10^{-6}$ & (0.00012, 0.00014, 0.00016) \\
\hline
\end{tabular}
\end{center}
\end{table}

While many of the parameters do not lend well to the traditional $H_0: \theta = 0$ hypothesis testing as they must be positive.  We can conduct simple hypothesis tests on the $\alpha$ parameters to look for significant changes due to interventions using contrasts.  Specifically the sequential contrasts of $\alpha_1 -\alpha_0,~\alpha_2-\alpha_1,~\alpha_3 -\alpha_2,~\alpha_4 -\alpha_3$ and $\alpha_5 - \alpha_4$.  These contrasts quantify the changes that in transmission rate from Susceptible to Exposed due to the interventions and are what policy makers want to see.  Furthermore, they want a statistical test on whether or not the intervention performed in a statistically significant manner.  This can be done by simply subtracting the MCMC samples to generate the contrast of interest.  Using these subtracted samples one can look at the mean, standard deviation, quantiles and the proportion of samples above 0, $P(>0)$.  Table~\ref{tbl:ConEst} shows these quantities for the contrasts listed above.  Notice that the intervention at day 12 reduced the transmission rate by approximately $4.45\times 10^{-7}$ which is considerable and the proportion of samples above 0 was $0.000$ indicating a statistically significant change due to the intervention.  The intervention taken at day 24, $\alpha_2-\alpha_1$, actually increased the transmission rate, where the intervention taken at day 28, $\alpha_3-\alpha_2$, then reduced the transmission rate.  Similarly the other two interventions increased and then decreased the transmission rates, respectively.  Furthermore, all interventions be deemed statistically significant since $P(>0)$ is either 0.000 or 1.000 indicating significance.

\begin{table}[ht!]
\caption{Mean, Standard Deviation, ($Q_{0.025},Q_{0.5},Q_{0.975}$) and proportion of samples larger than zero $P(>0)$ for sequential contrasts across $\bm{\alpha}$. Based on 5,000 samples from the posterior distribution}\label{tbl:ConEst}
\begin{center}
\begin{tabular}{l c c c c} \hline
Contrast & Mean & Std Dev. & ($Q_{0.025},Q_{0.5},Q_{0.975}$) & $P(>0)$ \\
\hline
$\alpha_1 - \alpha_0$ & -4.45$\times 10^{-7}$ & 2.81$\times 10^{-10}$ & ( -4.450$\times 10^{-7}$, -4.456$\times 10^{-7}$, -4.462$\times 10^{-7}$) & 0.000 \\
$\alpha_2 - \alpha_1$ & 4.04$\times 10^{-7}$ & 7.59$\times 10^{-10}$ & ( 4.02$\times 10^{-7}$, 4.04$\times 10^{-7}$, 4.05$\times 10^{-7}$) & 1.000 \\
$\alpha_3 - \alpha_2$ & -3.66$\times 10^{-7}$ & 2.05$\times 10^{-9}$ & ( -3.70$\times 10^{-7}$, -3.66$\times 10^{-7}$, -3.63$\times 10^{-7}$) & 0.000 \\
$\alpha_4 - \alpha_3$ & 1.76$\times 10^{-7}$ & 2.66$\times 10^{-9}$ & ( 1.71$\times 10^{-7}$, 1.76$\times 10^{-7}$, 1.80$\times 10^{-7}$) & 1.000 \\
$\alpha_5 - \alpha_4$ & -4.08$\times 10^{-8}$ & 2.11$\times 10^{-9}$ & ( -4.47$\times 10^{-7}$, -4.11$\times 10^{-7}$, -3.59$\times 10^{-7}$) & 0.000 \\
\hline
\end{tabular}
\end{center}
\end{table}

The model formulation also allows for the individual transmission rates to be computed by simply summing up the $\alpha_k$ through to the desired time point.  Table~\ref{tbl:AlphaEst} gives the mean, standard deviation and ($Q_{0.025},Q_{0.5},Q_{0.975}$) for the transmission rate of Exposed to Infected across each time interval.  This is done by simply add the corresponding MCMC samples.  This is another perspective on how the transmission rate changes across the time frame.  Notice that all of the transmission rates are positive which is required by the model specification.  Also notice that the mean transmission rates vary in orders of magnitude from $2.53\times10^{-9}$ to $2.33\times10^{-7}$. One interesting point that should be made is the highest transmission rate is at the beginning and the lowest transmission rate is at the end.  This is evidence that the interventions that the Qatari government has ultimately reduced the transmission rate.

\begin{table}[ht!]
\caption{Mean, Standard Deviation and ($Q_{0.025},Q_{0.5},Q_{0.975}$) of transmission rate of Exposed to Infected for the specified time intervals.}\label{tbl:AlphaEst}
\begin{center}
\begin{tabular}{l c c c c} \hline
Time & Sum & Mean & Std Dev. & ($Q_{0.025},Q_{0.5},Q_{0.975}$) \\
\hline
$0\le t < 12$ &$\alpha_0$ & 2.33$\times 10^{-7}$ & 4.86$\times 10^{-10}$ &
(2.32$\times 10^{-7}$, 2.33$\times 10^{-7}$, 2.34$\times 10^{-7}$) \\

$12\le t < 24$ & $\sum_{k=0}^1 \alpha_k$ & 2.18$\times 10^{-8}$ & 9.15$\times 10^{-10}$ &
(2.01$\times 10^{-8}$, 2.18$\times 10^{-8}$, 2.29$\times 10^{-8}$) \\

$24\le t < 28$ & $\sum_{k=0}^2 \alpha_k$ & 2.14$\times 10^{-7}$ & 8.74$\times 10^{-10}$ &
(2.12$\times 10^{-7}$, 2.13$\times 10^{-7}$, 2.15$\times 10^{-7}$) \\

$28\le t < 40$ &$\sum_{k=0}^3 \alpha_k$ & 3.96$\times 10^{-8}$ & 1.27$\times 10^{-9}$ &
(3.78$\times 10^{-8}$, 3.93$\times 10^{-8}$, 4.19$\times 10^{-8}$) \\

$40\le t < 59$ & $\sum_{k=0}^4 \alpha_k$ & 4.15$\times 10^{-8}$ & 9.11$\times 10^{-10}$ &
(4.01$\times 10^{-8}$, 4.13$\times 10^{-8}$, 4.32$\times 10^{-8}$) \\

$59\le t $ & $\sum_{k=0}^5 \alpha_k$ & 2.53$\times 10^{-9}$ & 1.94$\times 10^{-9}$ &
(1.03$\times 10^{-10}$, 2.07$\times 10^{-9}$, 7.20$\times 10^{-9}$) \\
\hline
\end{tabular}
\end{center}
\end{table}

To assess the fit of the model the posterior predictive distribution was used and is given by:
\begin{eqnarray}
\pi( I_{new}(t), R_{new}(t), D_{new}(t) | \mathbf{D} ) &=& \int L( I_{new}(t), R_{new}(t), D_{new}(t)|\bm{ \alpha }, \beta_A, \beta,\gamma,\eta) \cr
&\times& \pi \left( \bm{ \alpha }, \beta_A, \beta, \gamma, \eta | \mathbf{D} \right) d \bm{ \alpha }d\beta_A d\beta d\gamma d \eta.
\end{eqnarray}
Using the samples 5,000 samples from the posterior distribution 5,000 samples were generated from the posterior predictive distribution.  At each time $t$ the median, 0.025 and 0.975 quantiles were obtained to form a posterior predictive interval.

\begin{figure}[ht!]
\begin{center}
  \begin{tabular}{ccc}
  (a) & (b) & (c) \cr
  \includegraphics[width = 0.3\textwidth]{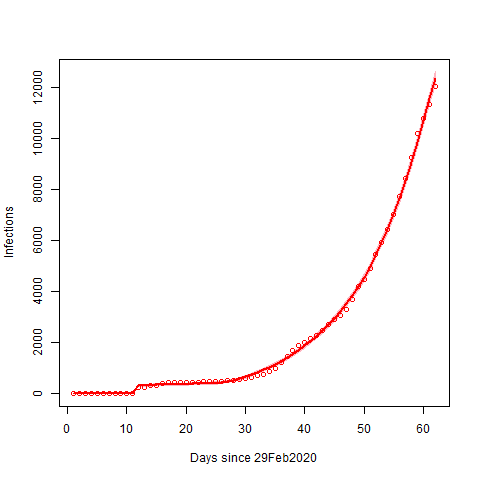} &
  \includegraphics[width = 0.3\textwidth]{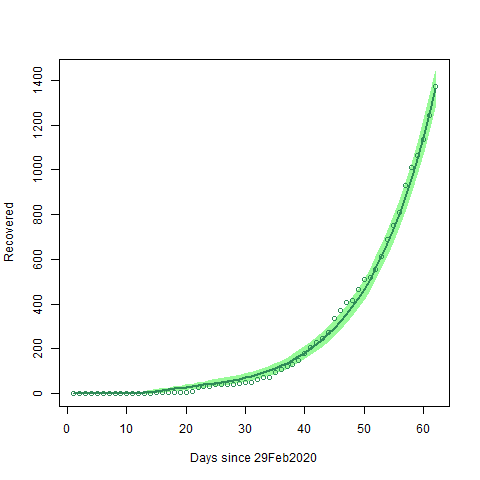} &
  \includegraphics[width = 0.3\textwidth]{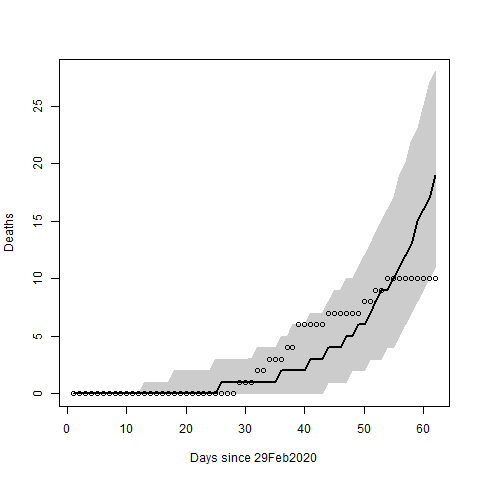}  \cr
  \end{tabular}\caption{Plots of Active Infections (a), Recovered (b) and Deaths (c) for the State of Qatar for the days since 29 February 2020 until 1 May 2020 with posterior predictive bands.  Posterior predictive bands are based on the 0.025 and 0.975 Quantiles from 5,000 samples from the posterior predictive distribution. \label{fig:PostData}}
\end{center}
\end{figure}

Figure~\ref{fig:PostData} shows the model fits for Active Infections, Recovered and Deaths with posterior predictive bands.  Notice that, the model does quite well at fitting the dynamics of the Active Infections including the jump at day 12 and captures  the plateau and the exponential growth after the plateau as well.  The Recovered model does fits well as does the Deaths data.  To assess the explained variance a pseudo-$R^2$ was formed using the median from the posterior predictive distribution at each time as the point estimates.  This resulted in a pseudo-$R^2$ of 0.998 which indicates the fitted model explains approximately 99.8\% of the variance in the data.  Based on this the model is deemed to fit well.  It should be noted that standard data splitting procedures for model validation are difficult in this scenario as removing values from the system may cause unstable behavior.

\section{Predictive Performance}\label{sec:PP}
To assess the model performance predictive performance is utilized with days from 2 May 2020 and 10 May 2020 used as a test set.  Using the samples from the posterior distribution the posterior predictive distribution was computed for each day of the test set and 95\% posterior predictive intervals were created using the 2.5\% and 97.5\% quantiles.  The test data is then compared to the posterior predictive intervals for each of the endpoints.

\begin{figure}[ht!]
\begin{center}
  \begin{tabular}{ccc}
  (a) & (b) & (c) \cr
  \includegraphics[width = 0.3\textwidth]{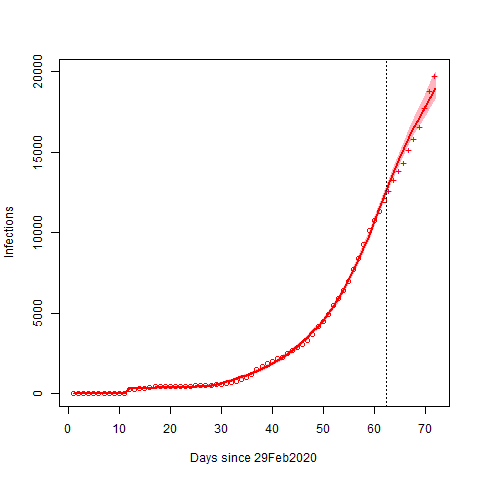} &
  \includegraphics[width = 0.3\textwidth]{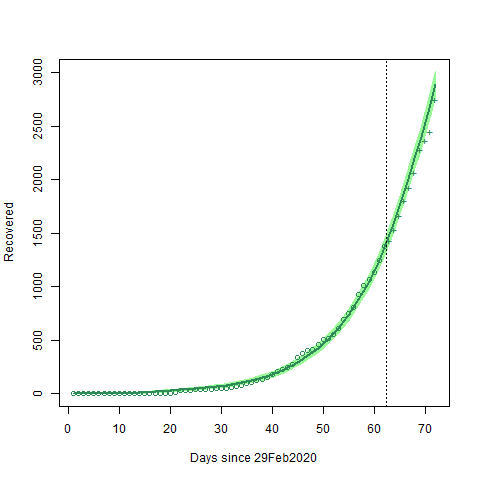} &
  \includegraphics[width = 0.3\textwidth]{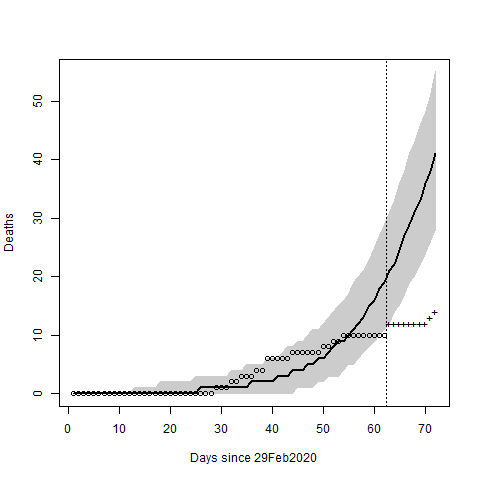}  \cr
  \end{tabular}\caption{Plots of Active Infections (a), Recovered (b) and Deaths (c) for Qatar for the days since 29 February 2020 until 1 May 2020 with posterior predictive bands with actual values from 2 May 2020 until 10 May 2020 denoted with (+).  Vertical line separates the training data from the predictions.  Posterior predictive bands are based on the 0.025 and 0.975 Quantiles from 5,000 samples from the posterior predictive distribution. \label{fig:PredData}}
\end{center}
\end{figure}

Figure~\ref{fig:PredData} shows the data, predictive bands and the actual values for Active Infections (a), Recovered (b) and Deaths (c).  Training data is presented by (o) and test set data is presented by (+) and the vertical line separates the training time frame from the prediction time frame.  Notice that, the Active Infections (panel a) performs incredibly well with all test set points in the predictive interval.  The Recovered (panel b) performs equally well.  Deaths (panel c) performs the worst as the model consistently over predicts the true number of deaths.  Overall the model does reasonably well at predicting the test set.

Another view of predictive performance is to examine pseudo-predictive-$R^2$ which compares the predicted values with the actual values for the test set.  This calculation leads to a pseudo-predictive-$R^2 \approx 0.989$ which is slightly lower than the pseudo-$R^2$ associated with the fit of the model to the data in the training set but is still very high.  Of course several other measures of predictive performance exist however this is the easiest to understand as it measures the amount of variation explained by the predictions across the test set.

\section{Discussion}\label{sec:Discussion}
This work has demonstrated how to build a SEIRD model for the Covid-19 outbreak in the State of Qatar, include interventions, estimate model parameters and generate posterior predictive intervals using a Bayesian framework.  Furthermore, the model is able to treat the Susceptible and Exposed compartment as latent variables, as no data is observed about them other than approximate initial values.  The model fits  the data quite well with a pseudo-$R^2 \approx 0.998$ and predicts reasonably well with pseudo-predictive-$R^2 \approx 0.989$.   One can also note that in the model definition, no immigration, emigration, natural births and natural mortality were not included and based on the high psuedo-$R^2$ would have a negligible effect on fit.  Furthermore, the model did not contain compartments for those who recovered without being confirmed infections.  As this was not observed one can only speculate on the impact that additional data would have on the model fit, however it would be very small.

\begin{figure}[ht!]
\begin{center}
  \includegraphics[width = 0.6\textwidth]{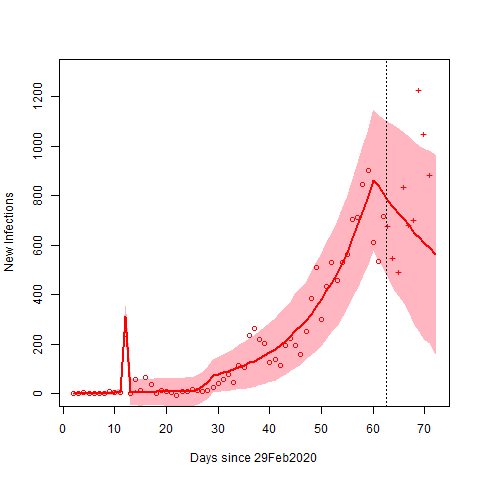} 
\caption{Plot of New Infections for Qatar for the days since 29 February 2020 until 1 May 2020 with posterior predictive bands with actual values from 2 May 2020 until 10 May 2020 denoted with (+).  Posterior predictive point estimates are based on the median and bands are based on the 0.025 and 0.975 Quantiles from 5,000 samples from the posterior predictive distribution at each time point. \label{fig:NewInfPredData}}
\end{center}
\end{figure}

The modeling paradigm is quite flexible for modeling the Covid-19 data as it easily incorporates interventions into the system and can quantify the impact of the intervention.  Furthermore, using simple differences the model can be used to predict new infections as well.  Figure~\ref{fig:NewInfPredData} shows the plots of the new infections with predictive bands based on 2.5\% and 97.5\% based on 5,000 samples from the posterior prediction distribution at each time point.  Notice that the model does well at capturing the jump at day 12 and the bands capture most of the data.  The drop beginning at the intervention at day 59 provides for the drop in daily infection rates.

\begin{figure}[ht!]
\begin{center}
  \begin{tabular}{c}
  \includegraphics[width = 0.6\textwidth]{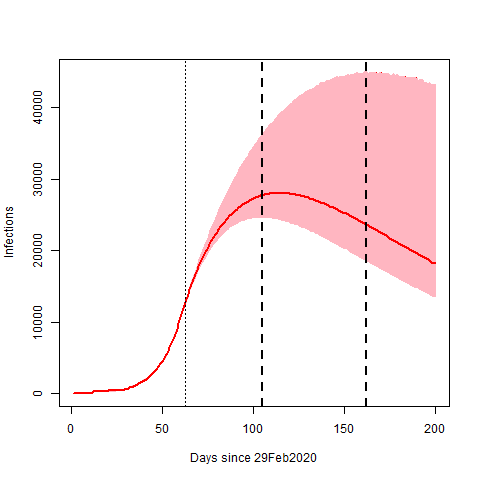}
  \end{tabular}\caption{Plot of Long-term New Infections for Qatar for the days since 29 February 2020 until 28 July 2020 with posterior predictive bands.  Training data from 29 February 2020 to 1 May 2020. Posterior predictive bands are based on the 0.025 and 0.975 Quantiles from 5,000 samples from the posterior predictive distribution.  Dark dashed vertical lines give the 95\% predictive interval for the maximum active infections. \label{fig:LongTerm}}
\end{center}
\end{figure}

Another use of the model may be for long term predictions.  While this is extrapolation it does provide policy makers a tool for planning, provided nothing changes, i.e. no interventions are taken.  It also allows policy makers to see the possible long term effects of their decisions.  Figure~\ref{fig:LongTerm} shows the long-term predictions of the model if no other interventions are made past 1 May 2020.  Notice that the predictions do eventually decrease across the future time frame.  Notice the width of the predictive bands for times farther in the future. This reflects the uncertainty associated with extrapolating into the future.  However, one item that we can calculate from this is a 95\% predictive interval for the peak infection time.  By simply recording the maximum value for each of the predictive distribution trajectories from the MCMC samples one can obtain a distribution of the time for the maximum.  In this case this gives the 95\% predictive interval for the maximum to be (105,162).  This means that the peak infection time will be between day 105 (13 June 2020) and day 162 (9 August 2020) of the outbreak given that no other interventions or process changes occur.  Fig.~\ref{fig:LongTerm} also shows this interval given by the dark dashed vertical lines.  The width of the interval quantifies the uncertainty about where the maximum active infections will occur.  Since the width of the interval is 57 days, this indicates that there is a large amount of uncertainty on when the number of active infections will begin to decline.

Future work could be to add an overdispersion parameter into the model to allow for the more accurate capture of uncertainty.  Furthermore, one can perform simulation studies to better understand how the model may perform under various scenarios.  Feature selection methods could be employed to select where the interventions should be placed as well as other forms of interventions could be included in the model.  Another possibility to address any deviations from the standard model a semi-parametric technique could be studied as well.

\section*{Acknowledgements}
The authors would like to acknowledge to the State of Qatar and the Ministry of Health for the daily updates and additional data.  In addition the authors would like thank Virginia Commonwealth University in Qatar and Qatar University for supporting this effort.


%
%
%
%

\end{document}